\documentstyle[12pt]{article}

\newlength{\dinwidth}
\newlength{\dinmargin}
\def\msb{\overline{\rm MS}}
\def\GeV{{\rm GeV}}

\def\lapproxeq{\lower .7ex\hbox{$\;\stackrel{\textstyle
<}{\sim}\;$}}
\def\gapproxeq{\lower .7ex\hbox{$\;\stackrel{\textstyle
>}{\sim}\;$}}

\def\be{\begin{equation}}
\def\ee{\end{equation}}
\def\bea{\begin{eqnarray}}
\def\eea{\end{eqnarray}}

\begin{document}
\titlepage
\begin{flushright}
RAL--TR--97--061\\
November 1997 \\
\end{flushright}
\begin{center}
\vspace*{2cm}
{\large \bf A Practical  Procedure for Evolving \\[2mm]
Heavy Flavour Structure Functions}\\
\vspace*{1cm}
R.\ S.\ Thorne$^*$ \\

{\it Theoretical Physics, University of Oxford,
1 Keble Road, Oxford, OX1 2NP}

and

R.\ G.\ Roberts \\

{\it Rutherford Appleton Laboratory, Chilton,
Didcot, Oxon, OX11 0QX}\\

\end{center}

\vspace*{1.5cm}
\begin{abstract}
The transition from the low $Q^2$ region in deep inelastic scattering
where charm production is described by photon-gluon fusion to a region
where the charm structure function $F_2^c$ is largely generated by a
charm quark density should be a smooth passage. 
The all orders prescription for matching between the two schemes does not 
uniquely determine all
the relevant coefficient functions (CF's) or guarantee the correct
threshold behaviour at fixed order in $\alpha_s$. The constraints of 
matching not only the value but also the {\it evolution}
of $F_2^c$ order by order allows {\it all} CF's to be determined at any 
order, each of which reduces to the appropriate massless $\msb$ expression 
as $Q^2 \rightarrow \infty$, 
and leads to a smooth transition in the threshold region.  This 
procedure gives an excellent description of the $F_2^c$ data for all 
$Q^2$.

\end{abstract}

\vspace*{3.0cm}

* Junior Research Fellow, Jesus College, Oxford.

\newpage

\noindent{\large \bf 1 {\hskip 0.5cm} Introduction}

\medskip

Recent measurements of charm production at HERA \cite{h1charm,zeuscharm}
 emphasise the
importance of having a consistent theoretical framework for heavy
flavour production in deep inelastic scattering (DIS). The charm
structure function $F_2^c$ could be 20\% or more of the total $F_2$
and a modern global analysis of structure functions must necessarily
include a satisfactory description of $F_2^c$. Recently 
\cite{bible} we have 
developed a general order-by-order procedure which solves the problem
of how to calculate $F_2^c$ in the region close to $Q^2=m_c^2$ where
$m_c$ is the charm quark mass. Below $Q^2=m_c^2$ the conventional
description in terms of order-by-order in $\alpha_s$ 
boson-gluon fusion is perfectly adequate,
 but for $Q^2 > m_c^2$ this description becomes increasingly unreliable
due to the presence of potentially large logarithms at all orders in 
$\alpha_s$ which should
be resummed.  By changing to the alternative description where the
charm quark is treated as a parton, at the point $Q^2=m^2$, this resummation
is automatically performed and, at the same time, a complete set of parton
densities needed to calculate other processes involving nucleons is 
obtained.
While this idea of transferring from one description to another is not
new, until now  there has not been a satisfactory procedure for preserving 
threshold behaviour across the transition point whilst systematically 
presenting the calculation to any well--defined order.

In this paper we present our results at leading order (LO) and 
next-to-leading order (NLO) to demonstrate the key features of our method,
the practical advantages it offers and how it improves upon previous
attempts to treat heavy flavours. We demonstrate that the 
requirement of all orders equivalence between the two descriptions does not 
actually provide unique expressions for every relevant CF.
The order-by-order in $\alpha_s$ matching
of the structure function itself between the two descriptions at the transition
point does not remove the ambiguity.
Extra information is required to do this and, in our case, this  is provided
by the further requirement that the order-by-order
{\it evolution} of the structure function should also be continuous
across the transition point. This added constraint then allows
each CF to be determined with the practical benefit that a
very smooth transition is ensured, the correct threshold behaviour being
maintained across the transition point and above.
Here we only sketch how our results are derived, the emphasis here being
on how the procedure works in practice; the detailed theoretical 
justification is presented in full in \cite{bible}.

\medskip

\noindent{\large \bf 2 {\hskip 0.5cm} The FFN, ZM-VFN and VFN schemes}

\medskip

Consider the case where charm is produced by the neutral current.
At low $Q^2$, the number of active flavours, $n_f$, is 3 and the charm
cross section is generated by photon-gluon fusion (PGF)
\cite{witten,grv}. This 
corresponds to the so-called fixed flavour number scheme 
(FFNS) and provides an acceptable description provided $Q^2$ is not
large and one is not interested in the concept of a charm quark density.

In the FFNS at order
${\cal O}(\alpha_s)$ the charm structure function is given by
\be
F_2^c(x,Q^2,m_c^2) \; = \;\frac{\alpha_s}{2\pi} 
\;C_g^{(1)\;FF}(Q^2/m_c^2) \:\otimes\:g_{n_f=3}(\mu^2),
\label{eq:lof2cffns}
\ee
where the CF which is convoluted with LO evolved gluon density 
$g_{n_f=3}$ is
\bea
 C^{(1)\;FF}_g (z,\epsilon) & = &
\biggl [ \;(P^0_{qg}(z)
+ 4\epsilon z (1-3z) -8\epsilon^2 z^2)\; \ln  \biggl ( \frac{1+v}{1-v}
\biggr ) \nonumber \\
 & + &  (8z(1-z)-1-4\epsilon z(1-z))v \biggr ] \; \theta(\hat W^2-4m_c^2) 
\label{eq:lochgffns}
\eea
where $\epsilon=m_c^2/Q^2$, $\hat W^2=Q^2(1/z-1)$, the gluon quark centre 
of mass energy, $v$ is the velocity of the charm quark or antiquark in the 
photon--gluon centre--of--mass frame, defined by $v^2=1-4m_c^2/\hat W^2$,
and $P^0_{qg}(z) = z^2 + (1-z)^2$, the LO quark-gluon splitting function. 
 These $v$--dependent terms ensure 
that the coefficient function tends to zero
smoothly as $\hat W^2=4m_c^2$ is approached from below,
and hence the structure function has a smooth threshold in 
$W^2=Q^2(1/x-1)$. 

In the limit $Q^2\to \infty$ the gluon CF reduces to
\be
 C^{(1)\;FF}_g (z,\epsilon) \; \rightarrow \; P^0_{qg}(z)\ln(\epsilon)+
P^0_{qg}(z)\ln\biggl(
\frac{(1-z)}{z}\biggr) + 8z(1-z)-1
\label{eq:limita}
\ee
from which we can see the potentially large logarithm $\ln \epsilon$ =
$\ln (Q^2/m_c^2)$. The natural way to sum this contribution together with 
higher powers of such logs is to describe charm production through
a charm density which evolves according to the standard (light quark)
renormalisation group equations.   

This leads us onto an alternative description which is simple (and
therefore practical) and which is commonly used in global analyses of DIS
data \cite{cteq,mrs} where the aim is to extract parton densities 
for both light and heavy flavours.
In this scheme, which we label the zero mass variable flavour scheme
(ZM-VFNS), a value of $\mu_c^2$ ($\sim m_c^2$) is chosen below which
$m_c$ is taken to be infinite and above which $m_c$ is taken to be zero. 
While this is intended to be a rough description only it has proved
successful in describing the then
existing (pre-HERA) charm structure function data well 
above the threshold region with $\mu_c^2\approx 3\GeV^2$.
However it is bound to be unrealistic at low $Q^2$ because of the abrupt
variation in $Q^2$ -- in contrast to the proper smooth threshold behaviour
in $W^2$.  As $Q^2$ increases the ZM-VFNS becomes more reliable, the
${\cal O}(\alpha_s)$, i.e. the NLO expression  in this case, being given by
\be
F_2^c(x,Q^2,m_c^2) = C_c^{n_f=4}(Q^2/\mu^2)\:\otimes\:c_+(\mu^2,m_c^2/\mu^2)
 \:\:+\:\: C_g^{n_f=4}(Q^2/\mu^2)\:\otimes\: g_{n_f=4}(\mu^2,m_c^2/\mu^2)
\label{eq:f2czmvfns}
\ee
where the charm density $c_+ \equiv \bar c + c$ and 
the CF's are expanded as
\bea
C_c^{n_f=4}(z,1) &=& C_c^{(0)}(z) + \frac{\alpha_s}{2\pi} \; 
C_c^{(1)\;\msb}(z)\nonumber \\
C_g^{n_f=4}(z,1) &=& \qquad \qquad \;\frac{\alpha_s}{2\pi} \;
C_g^{(1)\;\msb}(z)
\label{eq:cgzmvfns}
\eea
where
\be
C_c^{(0)}(z)=z\;\delta(1-z) \:\:\:\: \mbox{and}\:\:\:\:
C_g^{(1)\;\msb}=P_{qg}^0(z)\;\ln \biggl (\frac{1-z}{z} \biggr )
+8z(1-z) -1. 
\label{eq:cgmsbar}
\ee

Our aim is to develop an approach which extrapolates smoothly from the
FFNS at low $Q^2$ to the ZM-VFNS at high $Q^2$, maintaining the correct 
ordering in both schemes. First we note that 
in the FFNS eq(\ref{eq:lof2cffns}) is valid up to corrections of
${\cal O}(\Lambda^2/m_c^2)$ while in the ZM-VFNS eq(\ref{eq:f2czmvfns})
is valid only up to corrections of ${\cal O}(m_c^2/\mu^2)$. In order 
to improve the accuracy of eq(\ref{eq:f2czmvfns}) we need to
examine the connection between the parton densities in the two schemes.
Eq(\ref{eq:f2czmvfns}) actually has an apparent extra degree of freedom over 
eq(\ref{eq:lof2cffns}), namely the charm density evaluated at its
starting scale. This is of course not really true in practice and the
connection between the parton densities for 3 and 4 flavours takes the form 
\bea
c_+(z,\mu^2,m_c^2/\mu^2) &=& A^{cg}(\mu^2/m_c^2) \:\otimes\: 
g_{n_f=3}(\mu^2) \nonumber \\
g_{n_f=4}(z,\mu^2,m_c^2/\mu^2) &=& A^{gg}(\mu^2/m_c^2) \:\otimes\: 
g_{n_f=3}(\mu^2) \label{eq:partons}
\eea 
at leading order, 
where the elements $A^{ba}$ which contain $\ln(\mu^2/m_c^2)$ terms,
are, in general, part of a full 5$\times$4
matrix which also connects the light quark flavours. 

For $Q^2 \gg m_c^2$, the equivalence of the FFNS and the ZM--VFNS at all 
orders then lead to the connections between the CF's in 
the two schemes
up to ${\cal O}(m_c^2/\mu^2)$ \cite{buza}, in particular up to 
${\cal O}(\alpha_s^2)$
\bea
C_g^{FF}(z,Q^2/\mu^2,Q^2/m_c^2) &=& C_c^{n_f=4}(Q^2/\mu^2) \:\otimes\:
A^{cg}(\mu^2/m_c^2) \nonumber \\
 &\:&+\:\: C_g^{n_f=4}(Q^2/\mu^2) \:\otimes\:
A^{gg}(\mu^2/m_c^2) \:+\: {\cal O}(m_c^2/\mu^2). 
\label{eq:cfconn}
\eea
The details of the connection are fully worked out in \cite{buza}. 
To improve
the accuracy of eq(\ref{eq:cfconn}), where the uncertainty is reduced
to ${\cal O}(\Lambda^2/m_c^2)$, requires defining `corrected' CF's, 
$C_b^{VF}$, in another $n_f=4$ scheme 
-- the variable flavour number scheme (VFNS) --
where one can write
\bea
F_2^c(x,Q^2,m_c^2) &=& C_c^{VF}(Q^2/\mu^2,m_c^2/\mu^2) \:\otimes\:
c_+(\mu^2,m_c^2/\mu^2) \nonumber \\
 &\:&+\:\: C_g^{VF}(Q^2/\mu^2,m_c^2/\mu^2)  
\:\otimes\: g_{n_f=4}(\mu^2,m_c^2/\mu^2) \:+\: {\cal O}(\Lambda^2/m_c^2),
\label{eq:facttheha}
\eea
where the corrected CF's are related to the FFNS CF's by
\be 
C_b^{VF}(z,Q^2/\mu^2,m_c^2/\mu^2) = C_a^{FF}(Q^2/\mu^2,m_c^2/\mu^2)
\:\otimes\: \biggl [ A^{ba}(\mu^2/m_c^2) \biggr ]^{-1},
\label{eq:corrcoeff}
\ee
the new $n_f$=4 CF's now being {\it exact} at all values of $Q^2$.

In the spirit of ref\cite{cwz} the procedure we adopt is to use the FFNS
for $Q^2 \leq m_c^2$ and to switch to the VFNS for $Q^2 \geq m_c^2$.
The precise choice of the transition point is undetermined, though is 
naturally of order $m_c^2$. 
However, there are matching conditions between the partons in the two schemes
in order to ensure consistency with the usual asymptotic expressions,
e.g. eq(7), and 
taking $\mu^2=m_c^2$ removes complications arising
from $\ln(\mu^2/m_c^2)$ terms in these matching conditions.
To make practical headway in the VFNS
one must solve eq(\ref{eq:corrcoeff}) for the $C_b^{VF}$. However one can
see that the all orders matching of $F_2^c$ in the two schemes, from which 
eq(\ref{eq:corrcoeff}) arose, is not sufficient since, for example,
at low orders  the {\it single} 
quantity $C_g^{FF}$ is expressed in terms of the {\it two} quantities
$C_c^{VF}$ and $C_g^{VF}$. We stress that any choice satisfying eq(8) is 
``correct'' in the sense that it leads to the same all orders expression. 
Nevertheless,
each choice leads to a different expression if one uses the usual rules of 
combining coefficient functions and parton distributions of a given order to
obtain a fixed order in $\alpha_s$ expression for the structure function.
In principle a completely consistent theoretical approach would lead to a 
completely seamless transition between the two schemes at every order in 
perturbation theory. In practice this seems to be, at the least, incredibly 
difficult and is probably not even possible in any unique manner. However, in 
order to obtain as well--defined a theoretical procedure as possible we 
impose not only continuity of the structure function 
but also demand, in addition, order-by-order 
matching of the evolution of $F_2^c$ in the two schemes, as we shall see below.

\medskip

\noindent{\large \bf 3 {\hskip 0.5cm} The practical solution}

\medskip

The explicit form of eq({\ref{eq:partons}) for $\mu^2=Q^2$ at
${\cal O}(\alpha_s)$ is
\bea
c_+(z,Q^2) &=& \frac{\alpha_s}{2\pi}\;
\ln \biggl (\frac{Q^2}{m_c^2} \biggr )
\;P_{qg}^0 \:\otimes\: g_{n_f=3} \nonumber \\ 
g_{n_f=4}(z,Q^2) &=& g_{n_f=3}(z,Q^2) \:-\:\frac{\alpha_s}{6\pi}
\;\ln \biggl (\frac{Q^2}{m_c^2} \biggr )\; g_{n_f=3}. 
\label{eq:partdefcharm}
\eea
Inserting the implied expressions for the matrix elements 
$A^{cg}(z,\mu^2/m_c^2)$ and $A^{gg}(z,\mu^2/m_c^2)$
into eq(8) gives the familiar relation
\be
C_g^{(1)\;FF}(z,Q^2/m_c^2) = C_g^{(1)\;VF}(z,Q^2/m_c^2) 
\:+\: C_c^{(0)\;VF}(Q^2/m_c^2) \:\otimes\: P_{qg}^0\;
\ln \biggl (\frac{Q^2}{m_c^2} \biggr )
\label{eq:cfgconnect}
\ee
connecting the gluonic CF's in the FFNS and VFNS. In previous 
implementations of a VFNS \cite{acot}, eq(\ref{eq:cfgconnect}) served
as the definition for $C_g^{(1)\;VF}$ in terms of the PGF CF 
(eq(\ref{eq:lochgffns})) with an assumed form of $C_c^{(0)\;VF}$
given by
\be
\hat C_c^{(0)\;VF}(z,Q^2/M^2)= z\;\delta(\hat x_0-z)\biggl(1+
{4m_c^2\over Q^2}\biggr), \hskip 0.4in \hat x_0=\biggl(1+{m_c^2\over Q^2}
\biggr)^{-1}
\label{eq:acota}
\ee
where the delta-function describes the tree-level diagram for
a massive quark scattering from a photon
and the modified argument of the delta-function follows from 
demanding that the massive quark is on-shell. We note that the same 
definition of the zeroth order coefficient function is adopted in 
\cite{mrrs}
though there are differences between this and \cite{acot}, notably a 
mass dependent evolution.
  
We do not believe that it is appropriate to go suddenly from a scheme in 
which there is no charm parton distribution to one where the evolution and 
zeroth order coefficient function are such that the charm parton
behaves much like any other parton, but that the zeroth order coefficient 
function should reflect the true physics near threshold.
Indeed, there are several reasons
why the above prescription is unsatisfactory. Usually at leading order in the 
expressions for structure functions involving only light quarks we 
include just the zeroth order coefficient function, and thus the expression
is the summation of $\alpha_s(Q^2)\ln(Q^2)$ terms only. Combining 
both $C_c^{(0)\;VF}(Q^2/m_c^2)$ and $C_g^{(1)\;VF}(z,Q^2/m_c^2)$ 
with lowest order parton distributions we see that only the former 
really contributes to the LO expression, the first being down by a power of 
$\ln(Q^2)$ at large $Q^2$. However, keeping the simple 
form of $\hat C_c^{(0)\;VF}$ above leads to an abrupt rise of the charm
structure function as soon as $Q^2 > m_c^2$, which is totally incompatible 
with the smooth behaviour in $W^2$ required physically. Thus, previous
implementations \cite{acot} of a VFNS have included both 
both $C_c^{(0)\;VF}(Q^2/m_c^2)$ and $C_g^{(1)\;VF}(z,Q^2/m_c^2)$ in their LO
definition of the heavy quark structure function, and improved the smoothness 
by also changing the scale from $\mu^2=Q^2$ to something like  
$\mu^2=m_c^2+0.5\;Q^2\;(1-m_c^2/Q^2)^2$. Thus, we seem to have the choice of 
either a complete lack of smoothness in the charm structure function near 
threshold or an unconventional, and strictly incorrect ordering of the 
expression. We also have the disquieting feature that none of the heavy 
quark CF's contain the correct threshold behaviour in $W^2$
for $F_2^c$. The correct threshold behavior can only be obtained overall by 
a cancellation of sometimes large incorrect contributions between 
different terms. All this just emphasises that the above procedure is not
sufficient to yield a unique expression for the structure function in
the VFNS.

In order to better reflect the true physics in our choice of 
coefficient functions let us now consider the evolution of $F_2^c$. 
From eq(\ref{eq:lof2cffns}) the 
LO expression in the FFNS for the $\ln Q^2$ derivative is simply
\be 
\frac{dF_2^c(x,Q^2,m_c^2)}{d\ln Q^2} = \frac{\alpha_s}{2\pi} \;
\frac{d C_g^{(1)\;FF}(Q^2/m_c^2)}{d\ln Q^2} \:\otimes\: g_{n_f=3}(Q^2).
\label{eq:df2ffns}  
\ee
The corresponding expression obtained by differentiating the LO expression  
in the VFNS, for $Q^2$ just above $m_c^2$, is
\bea 
\frac{dF_2^c(x,Q^2,m_c^2)}{d\ln Q^2} &=& 
\frac{dC_c^{(0)\;VF}(Q^2/m_c^2)}{d\ln Q^2}\:\otimes\: c_+(Q^2) \nonumber \\
&+& \frac{\alpha_s}{2\pi}\;C_c^{(0)\;VF}(Q^2/m_c^2)\:\otimes\:
\biggl ( P_{qg}^0 \;\otimes\; g_{n_f=4}(Q^2) \:+\:
P_{qq}^0 \;\otimes\; c_+(Q^2) \biggr ).
\label{eq:df2vfns}
\eea
Throughout we are dropping the $n_f$ label on $\alpha_s$ to simplify
the notation. The question of how to treat the change in $\alpha_s$ 
from $n_f=3$  to
$n_f=4$ is discussed in our long paper \cite{bible}. Now for $Q^2=m_c^2$, the
terms in eq(\ref{eq:df2vfns}) involving $c_+$ vanish because of
eq(\ref{eq:partdefcharm}) and so demanding continuity of the evolution
across the transition point immediately leads, from 
eqs(\ref{eq:df2ffns},\ref{eq:df2vfns}), to 
\be
C_c^{(0)\;VF}(Q^2/m_c^2) \:\otimes\: P_{qg}^0 = 
\frac{dC_g^{(1)\;FF}(z,Q^2/m_c^2)}{d\ln Q^2}.
\label{eq:deflocf}
\ee
Generalising this relation to be the definition of 
$C_c^{(0)\;VF}(z,Q^2/m_c^2)$
at {\it all} $Q^2$ offers a series of real advantages. Firstly, from
the phenomenological point of view, it guarantees a smooth passage 
for the charm structure function from 
$Q^2 < m_c^2$ to $Q^2 > m_c^2$, by definition. Explicitly taking
the derivative of eq(\ref{eq:lochgffns}) is straightforward, giving
\bea
\frac{d C_g^{(1)\;FF}(z,\epsilon)}{d \ln Q^2} &=& 
\biggl[(P^0_{qg}(z)
+ 2\epsilon {z(1-2z^2)\over 1-z} -16\epsilon^2 z^2){1\over v} \nonumber \\
&+&(-4\epsilon z(1-3z) +16 \epsilon^2 z^2)\ln \biggl(\frac{1+v}{1-v}
\biggr) +(4\epsilon z(1-z))v\biggr] \theta(\hat W^2-4m_c^2)
\label{eq:lodchgffns}
\eea
and it is easy to see that in the limit $Q^2\to \infty$, 
\be
\frac{d C_g^{(1)\;FF}(z,\epsilon)}{d \ln Q^2} \rightarrow P^0_{qg}(z).
\label{eq:limitp}
\ee 
Hence, from eq(\ref{eq:deflocf}), we see that 
$C_c^{(0)\;VF}(z,\epsilon)$ must indeed
tend to the simple form $z\;\delta (1-z)$ in this limit, consistent 
with eq(\ref{eq:cgmsbar}). Also, since $C_g^{(1)\;FF}(z,\epsilon)$ 
contains the
factor $\theta(\hat W^2 - 4m_c^2)$ so does its $\ln Q^2$ derivative,
thus ensuring the correct threshold behaviour in $W^2$ for $C_c^{(0)\;VF}$
and in turn for $F_2^c$ at LO. Futhermore eq(\ref{eq:deflocf}) allows
the gluonic CF in the VFNS to be written as
\be
C_g^{(1)\;VF}(z,Q^2/m_c^2) = C_g^{(1)\;FF}(z,Q^2/m_c^2) \:-\:
\frac{dC_g^{(1)\;FF}(z,Q^2/m_c^2)}{d \ln Q^2} \; 
\ln \biggl ( \frac{Q^2}{m_c^2} \biggr ),
\label{eq:defnlogcf}
\ee 
and so now $C_g^{(1)\;VF}$ has the same threshold behaviour as $C_g^{(1)\;FF}$.
  
Finally, from eqs(\ref{eq:limita},\ref{eq:cfgconnect},\ref{eq:limitp}) 
we see that as $Q^2 \rightarrow \infty$,
$C_g^{(1)\;VF}(z,Q^2/m_c^2)$ does indeed tend to the correct asymptotic 
$\msb$ limit given by eq(\ref{eq:cgmsbar}). 

The extension of this procedure to any arbitrary order,
i.e. continuity of the derivative in the gluon sector, is described in
full in \cite{bible}. (Using our general approach continuity of the derivative in 
the singlet quark sector is not possible, but this has negligible effect in 
practice.)

The only price that may seem to have been paid in exchange for
obtaining a method which is far closer to true theoretical consistency
and a much improved phenomenological
description is the task of solving eq(\ref{eq:deflocf}) for the LO
charm CF, $C_c^{(0)\;VF}(z,Q^2/m_c^2)$. Note however that rather than
$C_c^{(0)\;VF}$ appearing explicitly in the second term of 
eq(\ref{eq:cfgconnect}), that term is now
replaced by the simpler term in eq(\ref{eq:defnlogcf}) and so we 
are just left with the term where $C_c^{(0)\;VF}$ is
convoluted with the charm density $c_+$. Calculating this expression
turns out to be easier than than deriving the form of 
$C_c^{(0)\;VF}$ itself, the convolution being given by
\bea
C_c^{(0)\;FF}(\epsilon)\otimes c_+(Q^2)
&=& - \int_{x}^{x_0} dz \, \frac{d C_g^{(1)\;FF}(z,\epsilon)} 
{d \ln Q^2}
\biggl({x\over z}\biggr)^2 \: \frac{d c_+(x/z,Q^2)}
{d (x/z)} \nonumber \\
&+& 3\int_{x}^{x_0} dz \, \frac{d C_g^{(1)\;FF}(z,\epsilon)} 
{d \ln Q^2}
\:{x\over z} \: c_+(x/z,Q^2) \nonumber \\
&-& 2\int_{x}^{x_0} dz \, \frac{d C_g^{(1)\;FF}(z,\epsilon)} 
{d \ln Q^2}
\int_{{x/z}}^1 d z'\, r(z') \: {x\over zz'}
\: c_+(x/zz',Q^2)
\label{eq:conv} 
\eea
where $x_0=(1+4\epsilon)^{-1}$ and $r(z)$ is given by
\be
r(z) = z^{1 \over 2}\biggl[ \cos\Bigl({\sqrt 7 \over 2}\ln 
{1\over z}\Bigr) + {3\over \sqrt 7} \sin \Bigl({\sqrt 7 \over 2}\ln 
{1\over z}\Bigr)\biggr]. 
\label{eq:defr}
\ee
 
Thus we can calculate the charm structure function at LO, taking
for $Q^2 < m_c^2$
\be
F_2^{c\;(0)\;FF}(x,Q^2,m_c^2) = \frac{\alpha_s}{2\pi}\:
C_g^{(1)\;FF}(Q^2/m_c^2)\:\otimes\: g_{n_f=3}(Q^2),
\label{eq:f20cffns} 
\ee
and for $Q^2 > m_c^2$
\be
F_2^{c\;(0)\;VF}(x,Q^2) = F_2^{c\;(0)\;FF}(x,m_c^2) \:+\:
C_c^{(0)\;VF}(Q^2/m_c^2) \:\otimes\: c_+(Q^2), 
\label{eq:f20cvfns}
\ee
where the constant term becomes almost negligible for $Q^2 > 4m_c^2$.
In the above, the partons are evolved only via the LO DGLAP equations.

For the NLO case, the partons are evolved via the full NLO equations.
The calculation of $F_2^{c\;FF}$ for $Q^2 < m_c^2$ adds in the 
${\cal O}(\alpha_s^2)$
contributions evaluated by Riemersma, Smith and van Neerven
\cite{rsvn}\footnote
{We are grateful to them for providing the program to compute these
corrections} and in our VFNS we include the contributions from
$C_g^{(1)\;VF}(Q^2/m_c^2) \otimes g_{n_f=4}$ and the extremely small 
contribution from $C_q^{(1)\;VF}(Q^2/m_c^2) \otimes \Sigma_{n_f=4}$, 
where the coefficient function in the latter is determined by continuity of 
the evolution in the gluon sector at NLO. 

We also briefly discuss the treatment of 
the longitudinal structure function $F_L^c$ since it is
an important example. The $\delta$-function
CF for $F_2^c$ of eq(\ref{eq:acota}), which previous procedures used,
contains a contribution from $F_L^c$ -- corresponding to the $4m_c^2/Q^2$
term. Thus in those descriptions, $F_L^c$ in the VFNS contains
a zero-order part, leading to a mis-match with the FFNS which starts
at order $\alpha_s$ with no $\ln(Q^2)$ enhancement. 
This implies rather odd $Q^2$ behaviour near threshold, see \cite{mrrs}.
In our approach we do not require this 
zeroth-order contribution to $F_L^c$ and so this problem
is avoided. Our prescription ensures a smooth 
transition as in the $F_2^c$ case, see \cite{bible}.

\medskip

\noindent{\large \bf 4 {\hskip 0.5cm} Results}  

\medskip

The results we present use a set of partons which are obtained from 
an analysis of DIS data in which charm and bottom
flavours are treated by the new procedure we have discussed here.
Only data on the full $F_2$ structure function are included in the fit,
not the data on $F_2^c$ though we shall compare with those data below.
The fits obtained are slightly sensitive to the value of $m_c$ and we find
$m_c \sim 1.35$ GeV gives the best fit. The value of $m_b$ was taken
to be 4.3 GeV. We have evidence already
that our more theoretically consistent treatment of heavy flavours improves  
the overall quality of the fits to $F_2$ \cite{bible}. 
We plan to incorporate
this procedure into a new (MRS type) global analysis where new data 
from hadronic collisions will be included and together provide updated
parton distributions.

In fig. 1 we show the result of the LO calculation 
of $F_2^c$ for two values of $x$. This demonstrates the success of 
our key result, namely the new form of the charm quark CF, which 
ensures a smooth transition from the description at low $Q^2$ in terms
of the FFNS to high $Q^2$ in terms of the ZM-VFNS. In contrast to
previous approaches \cite{acot} this has been achieved 
without resorting to an unusual
definition of LO structure functions (i.e. one in which the LO ZM--VFNS limit 
is not actually reached because ${\cal O}(\alpha_s)$ coefficients functions 
are included) or to any
cancellation between large individual contributions.

Fig.2 shows the similar situation for the NLO
case compared with the NLO ZM-VFNS result and the
continuation of the NLO FFNS expression. In the NLO case, the new
description remains close to the continuation of the FFNS result to 
larger $Q^2$ than in the LO case. This is what is expected since,
at the higher order, the FFNS includes more $\ln (Q^2/m_c^2)$ terms
which are resummed in the VFNS.

In fig.3 we show the comparison of the charm structure function
$F_2^c(x,Q^2)$ resulting from our NLO analysis with all available
data. The data at intermediate $x$ values comes from EMC \cite{emc}
measurements of inclusive muons, the new data from HERA is by
the H1 \cite{h1charm} and ZEUS \cite{zeuscharm} collaborations 
measuring $D$ and $D^*$ cross sections.
We show curves for three values of the charm mass and, as expected,
the sensitivity to $m_c$ is strongest at low $Q^2$. The values of
$F_2^c$ there from the EMC measurements suggest the higher value
$m_c$ = 1.5 GeV is favoured. Clearly, precise measurements of $F_2^c$ 
in this region would provide very accurate estimates of the charm mass.  

\medskip

\noindent{\large \bf 5 {\hskip 0.5cm} Summary}

\medskip

In this paper we draw attention to a new formulation
for describing heavy flavour structure functions in DIS. 
While the concept of the various schemes discussed in section 2
is not new, the realisation (discussed in section 3) that beyond all orders 
consistency and order-by--order
matching of the charm structure function across the transition from
the FFNS to the VFNS extra constraints are required to uniquely determine 
all the relevant CF's is new. In fact, previous attempts
to describe charm evolution suffered from both 
theoretical and phenomenological deficiencies. By demanding that
the order-by-order evolution of the charm structure function should be 
continuous as well as $F_2^c$ itself, we improve both the theoretical and 
phenomenological description of
charm (and bottom) production in DIS. 
This successful procedure, based on an examination of
the structure function expansion in a systematic order-by-order way, 
means that the procedure has a natural
generalisation to any given order and also to any given 
physical process. 

We have demonstrated that
the observed charm structure function is very well described
over a wide range of $x$ and $Q^2$ including especially the 
threshold region and that, consistent with the relatively large
importance of charm at low values of $x$,
fits to the total $F_2$ are improved. For full details of
our analysis for the general situation of heavy flavour
production in DIS we refer the reader to ref\cite{bible}.

\newpage

\noindent{\large \bf Figure captions}

\begin{itemize}
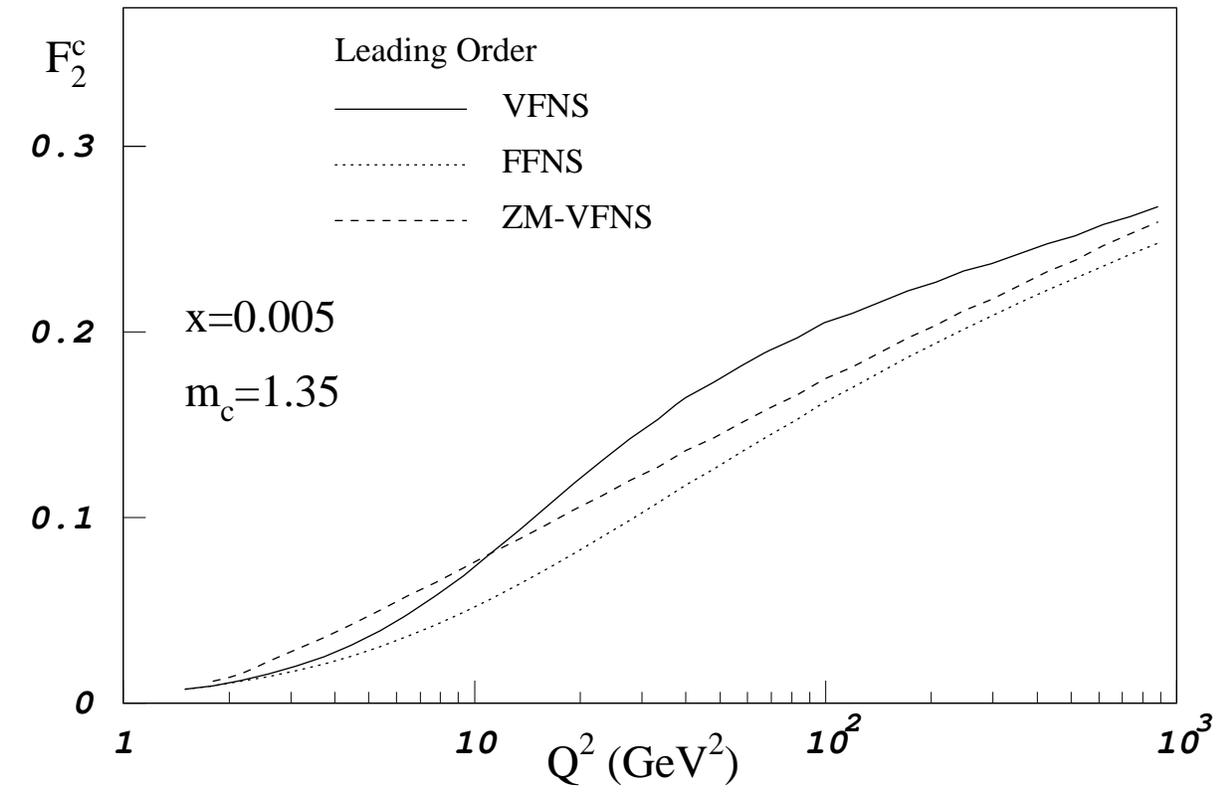
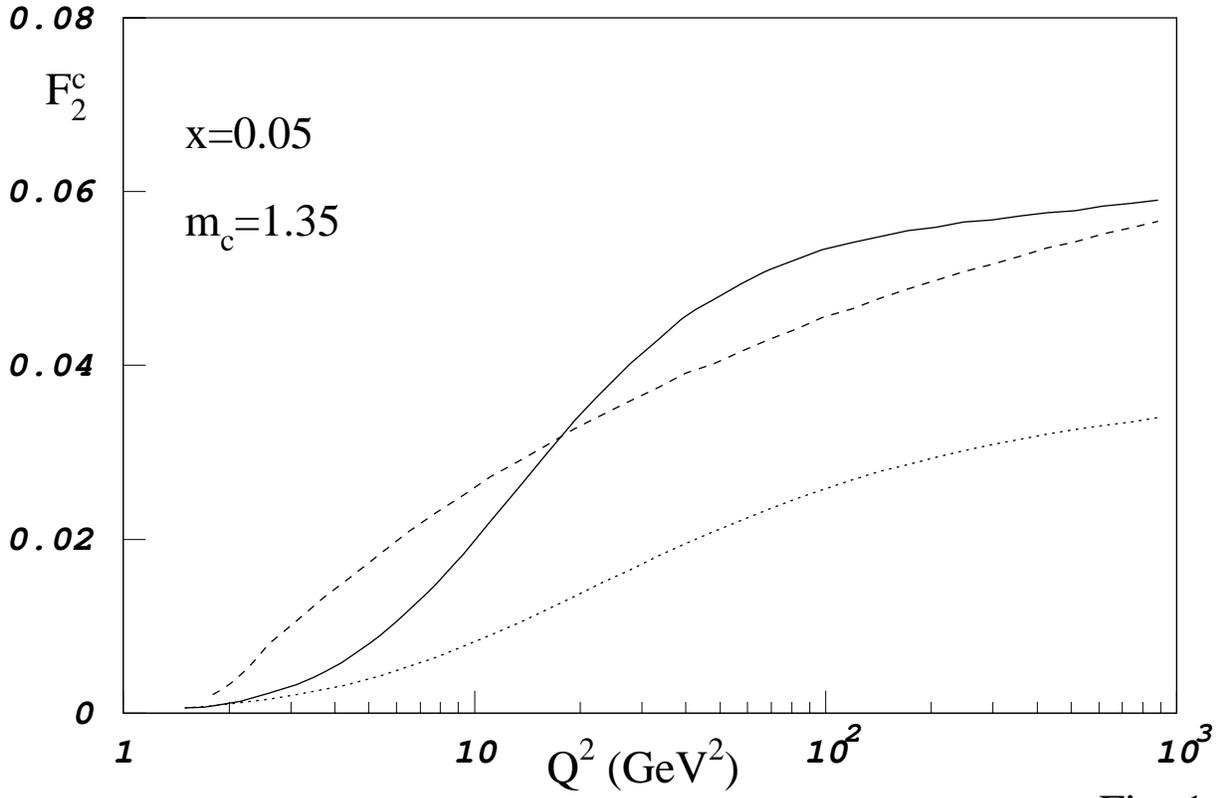
\item[Fig.~1]Charm quark 
structure function, $F_{2,c}(x,Q^2)$ for $x=0.05$ and 
$x=0.005$ calculated using our LO prescription, our input parton 
distributions evolved at LO and renormalization scale $\mu^2=Q^2$. Also shown 
are the continuation of the LO FFNS expression and the ZM--VFNS expression 
both calculated using the same parton distributions and same choice of scale.

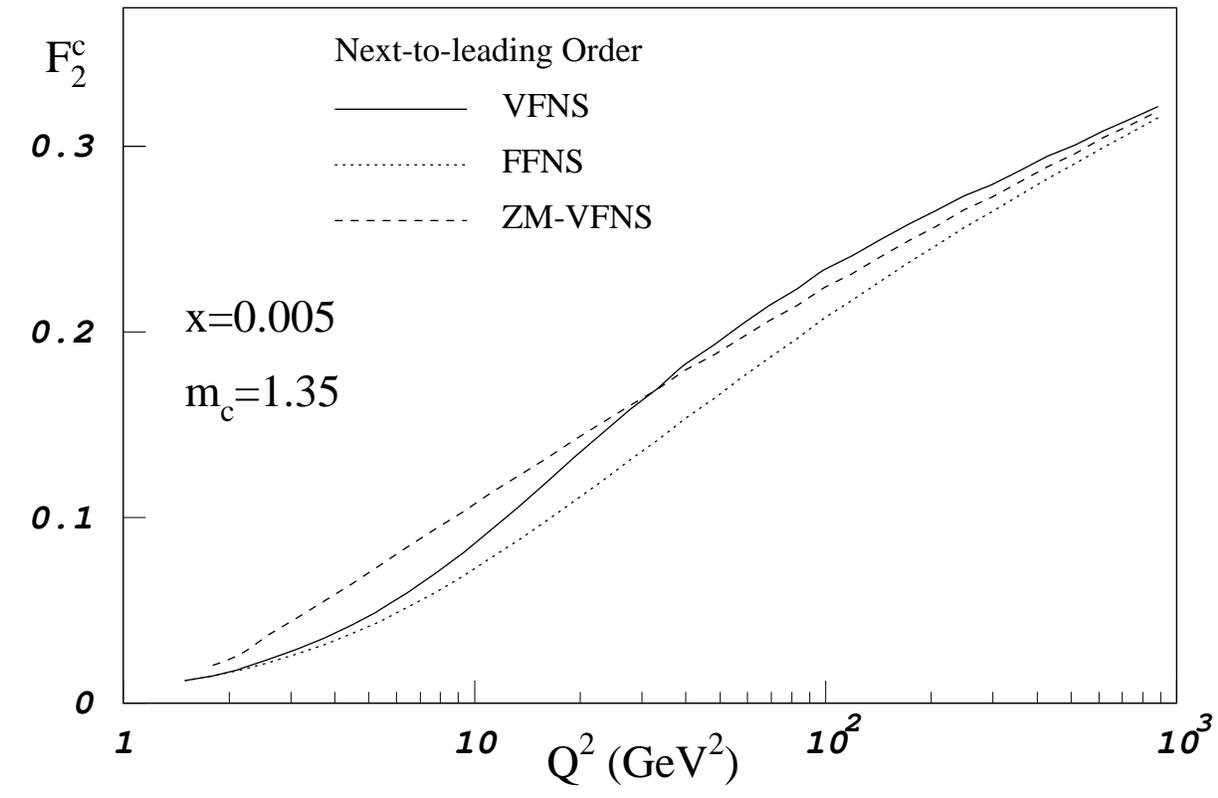
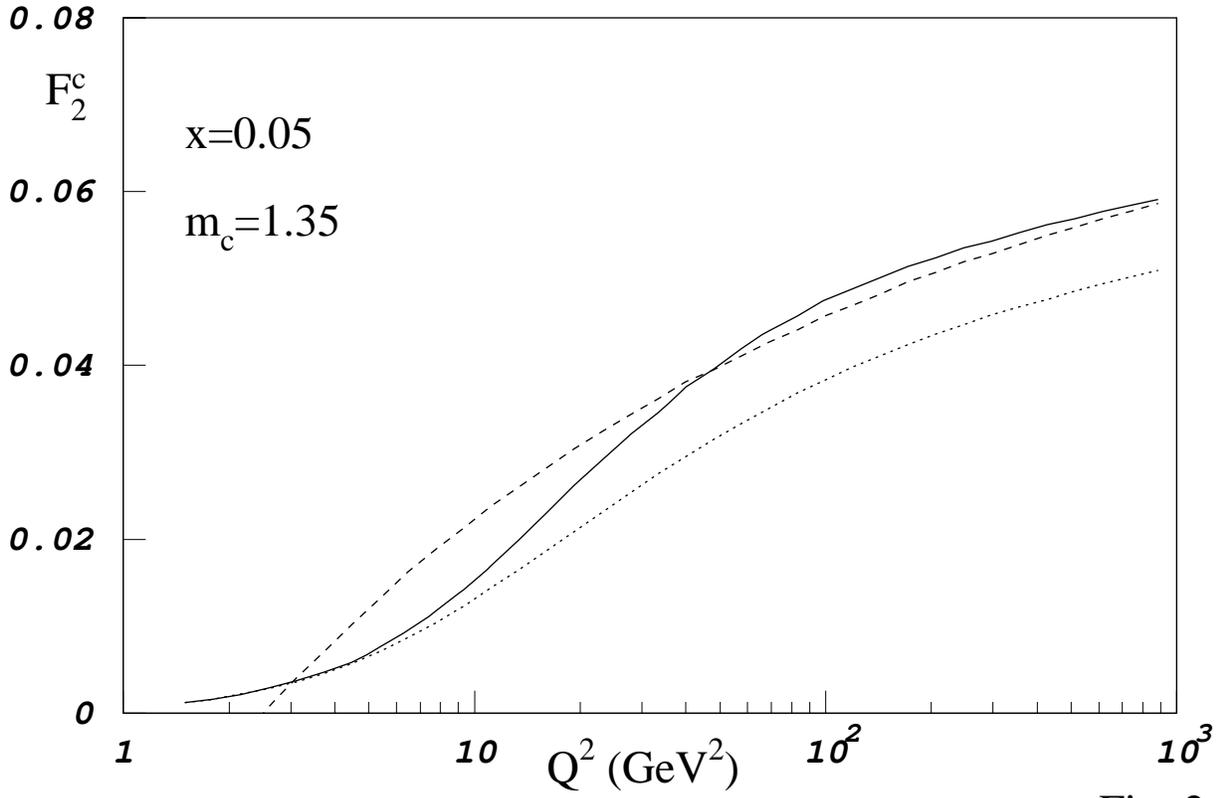
\item[Fig.~2]Same as fig.1 but with NLO prescriptions and NLO parton 
distributions.

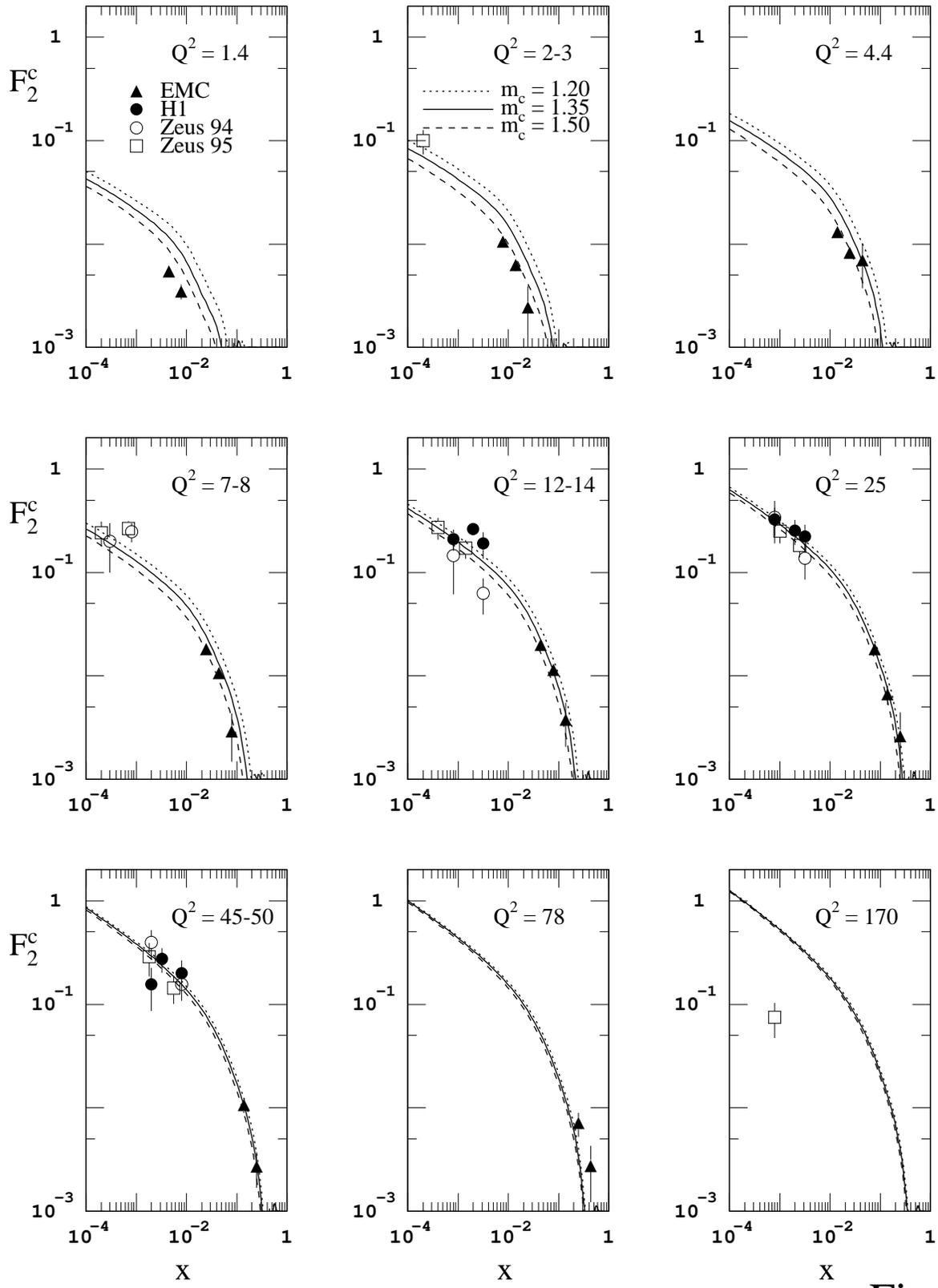
\item[Fig.~3]Our prediction for $F_2^c(x,Q^2)$ using our NLO prescription, 
the NLO partons obtained from our global fit and three different values of 
$m_c$ compared with data from EMC \cite{emc}, H1 \cite{h1charm},
ZEUS \cite{zeuscharm} (from the 1994 run) and preliminary data from 
ZEUS (1995 run) \cite{zeus95}.

\end{itemize}

\end{document}